\documentclass[twocolumn,pra,showpacs,superscriptaddress]{revtex4}
\usepackage{amssymb}
\usepackage{amsmath}
\usepackage{graphicx}
\usepackage{subfigure}
\usepackage{natbib}
\usepackage{epsfig}
\usepackage{amsfonts}
\usepackage{mathrsfs}
\usepackage{ulem}
\usepackage{color}
\usepackage[toc,page,title,titletoc,header]{appendix}
\usepackage{CJK}
\usepackage{graphicx}

\begin{document}

\title{$\mathcal{PT} $ symmetry breaking for the scattering problem in a one-dimensional non-Hermitian lattice model}

\author{Baogang Zhu}
\affiliation{Beijing National
Laboratory for Condensed Matter Physics, Institute of Physics,
Chinese Academy of Sciences, Beijing 100190, China}
\author{Rong L\"u}
\affiliation{Department of Physics, Tsinghua University, Beijing 100084, China}
\affiliation{Collaborative Innovation Center of Quantum Matter, Beijing, China}
\author{Shu Chen}
\email{schen@aphy.iphy.ac.cn} \affiliation{Beijing National
Laboratory for Condensed Matter Physics, Institute of Physics,
Chinese Academy of Sciences, Beijing 100190, China}
\affiliation{Collaborative Innovation Center of Quantum Matter, Beijing, China}

\begin{abstract}

We study the $\mathcal{PT} $-symmetry breaking for the scattering problem in a one-dimensional (1D) non-Hermitian tight-binding lattice model with balanced gain and loss distributed on two adjacent sites. In the scattering process the system undergoes a transition from the exact $\mathcal{PT} $-symmetry phase to the phase with spontaneously breaking $\mathcal{PT} $-symmetry as the amplitude of complex potentials increases. Using the S-matrix method, we derive an exact discriminant, which can be used to distinguish different symmetry phases, and analytically determine the exceptional point for the symmetry breaking. In the $\mathcal{PT} $-symmetry breaking region, we also confirm the appearance of the unique feature, i.e., the coherent perfect absorption Laser, in this simple non-Hermitian lattice model. The study of the scattering problem of such a simple model provides an additional way to unveil the physical effect of non-Hermitian $\mathcal{PT} $-symmetric potentials.

\end{abstract}



\maketitle
\date{today}

\section{Introduction}

It is known that one of the fundamental principles of quantum mechanics is that physical observables must be presented by Hermitian operators in the Hilbert space to guarantee real observables and probability conservation. However, in 1988 Bender and Boettcher found that a large class of non-Hermitian Hamiltonians can exhibit all real eigenvalues if these systems have parity-time ($\mathcal{PT} $) symmetry \cite{Bender}. The notable feature of these systems is that they undergo a transition from an exact $\mathcal{PT} $-symmetric (unbroken) phase to a spontaneously $\mathcal{PT} $-symmetry breaking phase, which is distinguished by the eigenvalues changing from all real to complex correspondingly. This leads to interesting phenomena in many fields, such as in quantum field theories and mathematical physics \cite{Bender04}, open quantum systems \cite{Rotter09}, Anderson models for disorder systems \cite{Goldsheid98,Heinrichs01,Molinari09}. The non-Hermitian systems have been experimentally realized in optical materials \cite{Regensburger13},
waveguide arrays \cite{Guo09,Ruter10},  microresonators \cite{WenJianming,LanYang14}, acoustic sensor \cite{Romain15,Zhangxiang2014} and LRC circuits \cite{Schindler11}.

Although whether the $\mathcal{PT} $ symmetric non-Hermitian Hamiltonian can define real quantum systems is not clear, efforts have been made to detect the effects of these non-Hermitian parts both in theoretical and experimental ways. Up to now, most of these studies rely on the basis that the scalar paraxial approximation of Maxwell's equation shares the same form with Schr\"odinger equation, in which the axial wave vector plays the role of energy \cite{ZinLin}. In order to find the original physical meaning of these non-Hermitian systems, a metric-operator theory is developed to map the non-Hermitian Hamiltonian to an equivalent Hermitian one \cite{Mosta04}. Based on these fundamental issues, the effects of the non-Hermitian terms have been investigated under different boundary conditions. Under the open boundary condition (OBC), the Bethe ansatz solution of a 1D tight-binding chain with two conjugated imaginary potentials at two end sites confirms the non-Hermitian $\mathcal{PT} $-symmetric quantum theory for the discrete systems \cite{Song09,Yuce15,Yogesh10}. Under the periodic boundary condition (PBC), the topological invariance is derived to find the topological properties in the quantum evolution of non-Hermitian models \cite{LiangShidong13}. The scattering propagation has also been investigated in various models, in which the scattering coefficients calculated by the transfer matrix method show the interesting unidirectional invisibility and self-emission effects \cite{Mostafa09,LiGe12,YDChong11,YDChong10,Mostafa13,Rotter13,Zhangxiang14,Longhi2010,Longhi10,Ahmed14}.

In this work we study the scattering propagation in the 1D tight-binding non-Hermitian lattice model whose balanced gain and loss are distributed on two adjacent sites to ensure the $\mathcal{PT} $-symmetry of the model. While the $\mathcal{PT} $-symmetry breaking in the eigenvalue problem of non-Hermitian lattice model has been extensively studied \cite{Song09,Yuce15,Yogesh10}, the $\mathcal{PT} $-symmetry breaking in the lattice scattering problem is not well understood yet.  The study of the simple lattice model can illustrate the most fundamental principle in the non-Hermitian scattering process and helps us to understand the more complicated phenomenon in other models. With the help of the transfer matrix $S $ we can analyze the stationary problem in the scattering propagation and define the form of the stationary wavefunction, thus getting the reflection and transmission coefficients. The $S $-matrix theory shows that the eigenvalues of the $S $ matrix remain unimodular in the exact $\mathcal{PT} $-symmetric phase whereas they have reciprocal moduli in the broken $\mathcal{PT} $-symmetric phase \cite{YDChong11}. The critical point is called the exceptional point (EP),  which signifies the breaking of the symmetry. In this paper we give the clear analytical results of eigenvalues of the $S $-matrix and the exceptional point for our studied model. We also derive a discriminant to find the EPs and observe the relationship between the real and complex potentials. Moreover, the significant property - coherent perfect absorption (CPA) Laser - can also be found in this simple model and the overall output coefficient confirms the results to be the typical CPA indeed. This model can be easily implemented on many experiments, such as optical material, waveguide arrays, and acoustic sensor.

The paper is organized as follows. In Sec. \uppercase\expandafter{\romannumeral2}, the model is presented. In Sec. \uppercase\expandafter{\romannumeral3}, we investigate the symmetry-breaking transition of the scattering process with the help of the $S $-matrix theory. In Sec. \uppercase\expandafter{\romannumeral4}, we focus on the CPA Laser and the overall output coefficient. A summary is given in Sec. \uppercase\expandafter{\romannumeral5}.

\section{Physical model}
We consider the simplest 1D $\mathcal{PT} $-symmetric model which consists of two lattice sites embedded with $\mathcal{PT} $-symmetric complex on-site chemical potentials in the infinite tight-binding linear main chain. As schematically shown in Fig. 1, the model can be described by the Hamiltonian:
\begin{eqnarray}
H &=&\sum J\hat{\phi} _{n+1}^{\dagger } \hat{\phi}_{n}+h.c. \nonumber\\
&&+(U+i\gamma )\hat{\phi} _{0}^{\dagger } \hat{\phi}_{0}+(U-i\gamma )\hat{\phi} _{1}^{\dagger } \hat{\phi}_{1}\label{Hamiltonian},
\end{eqnarray}
where $\hat{\phi}_{n}$ ($\hat{\phi}_{n}^{\dagger}$) denotes the annihilation (creation) operator for annihilating (creating) a mode state $|\phi_{n} \rangle $ at $n$th site. The main chain is an infinite isotropic tight-binding chain with $J $ being the nearest-neighbor hopping strength. The hopping terms provide the fluent channel of the continuum spectrum for the propagation of plane waves with dispersion $\omega = 2J\cos k $. The two defects are added with complex on-site chemical potentials, leading to the resonance at some income states during the propagation (as shown below). The real part of the potential $U $ is set to be $\left\vert U\right\vert < 2J $ to fulfill the resonance condition. For convenience, the hopping amplitude $J$ is set to be the unit of the energy ($J=1$).
\begin{figure}[!htb]
\includegraphics[width=8.5cm,]{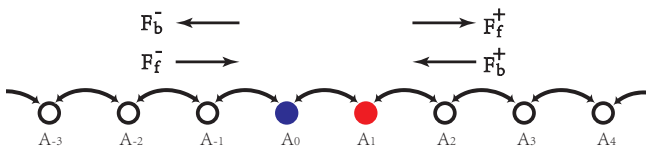}
\caption{(Color online) Schematic diagram of the model. $\mathcal{PT} $-symmetric complex on-site chemical potentials are embed on the two adjacent lattice sites in the infinite linear main chain. The couplings between different sites are indicated by the black arrows.}
\end{figure}

In general, $\mathcal{P}$ and $\mathcal{T}$ are defined as the space-reflection (parity) operator and the time-reversal operator, whose effects are $\mathcal{P}$: $p\rightarrow -p$, $x\rightarrow -x$\, and $\mathcal{T}$: $p\rightarrow -p$, $x\rightarrow x$, $i\rightarrow -i$. A Hamiltonian is said to be $\mathcal{PT} $ symmetric if it obeys the commutation relation $[\mathcal{PT},H]=0 $. In this discrete model the effect of $\mathcal{P} $ operator is $\mathcal{P} \hat{\phi}_{n}\mathcal{P}= \hat{\phi}_{-n+1}$ with the vertical line between site 0 and 1 as the mirror axis, and the effect of $\mathcal{T} $ operator is $\mathcal{T}i \mathcal{T}= -i $. So it's easy to prove that the Hamiltonian (\ref{Hamiltonian}) is invariant under the combined operation $\mathcal{PT} $.

\section{The scattering problem and $S $-matrix properties}

Using the $S $-matrix method, we now study the transport properties of the non-Hermitian $\mathcal{PT}$-symmetric system and analyze the $\mathcal{PT}$ symmetry breaking during the scattering process. If we expand the wave function of the system as $|\psi \rangle = \sum_{n} \phi_{n}(\tau) |\phi_n \rangle $ with $|\phi_n \rangle=\hat{\phi}_n^{\dagger} |0\rangle$, from the equation $i\partial_{\tau} |\psi \rangle = H |\psi \rangle$, we can derive the coupled-mode equation for expansion coefficients $\phi_{n}(\tau)$:
\begin{eqnarray*}
i\dot{\phi}_{n} &=&\phi _{n-1}+\phi _{n+1}+(U+i\gamma )\phi_{0}\delta_{n0}+(U-i\gamma )\phi_{1}\delta_{n1},
\end{eqnarray*}
where the overdot stands for the derivative of time $\tau$. The stationary solution can be expressed as the following form
\begin{equation*}
\phi _{n}(\tau)=A_{n}e^{-i\omega \tau },
\end{equation*}
and then we obtain the algebraic relationship of the amplitudes on each site:
\begin{eqnarray}
\omega A_{n}&=& A_{n-1}+A_{n+1}\nonumber\\
&+&(U+i\gamma )A_{0}\delta_{n0}+(U-i\gamma )A_{1}\delta_{n1}.
\end{eqnarray}
The above effective defect equation reveals that the wave propagates along the 1D chain with scattering center at sites 0 and 1 with the localized complex potentials.


For the scattering problem, the wave function can be expressed in the form
\begin{equation}
A_{n}=\left\{
\begin{array}{c}
F_{f}^{-}e^{ikn}+F_{b}^{-}e^{-ikn},n<0, \\
F_{f}^{+}e^{ikn}+F_{b}^{+}e^{-ikn},n>1,
\end{array}
\right.
\end{equation}
where $F_{f(b)}^{+(-)}$ stands for the coefficient of the two compositions in the whole wave function, the superscript $- $ stands for the left side of the scattering center whereas $+ $ the right side of the scattering center, and $f $ stands for the forward-going composition whereas $b $ the backward-going composition. $F_{f}^{-} $ is the so-called left-injection coefficient and $F_{b}^{+}$ is the right-injection coefficient. The momentum $k $ is assumed to be positive, and it is related to the incoming wave frequency $\omega $ by the dispersion equation $\omega=2 \cos k$. By substituting the expression $A_n$ into the effective defect equation (2) and applying the $S $-matrix method, we obtain the expression of all the coefficients in the wave function as
\begin{equation}
\left(
\begin{array}{c}
F_{b}^{-} \\
F_{f}^{+}%
\end{array}%
\right) =S\left(
\begin{array}{c}
F_{b}^{+} \\
F_{f}^{-}%
\end{array}%
\right) ,
\end{equation}
with the elements of $S$-matrix as
\begin{eqnarray}
S_{11} &=&-2i\sin ke^{-ik}/\Gamma ,  \notag \\
S_{12} &=&[-U^{2}-\gamma ^{2}+2\gamma \sin k+2U\cos k]/\Gamma ,  \notag \\
S_{21} &=&[-U^{2}-\gamma ^{2}-2\gamma \sin k+2U\cos k]e^{-2ik}/\Gamma ,
\notag \\
S_{22} &=&-2i\sin ke^{-ik}/\Gamma ,  \label{Smatrix}
\end{eqnarray}%
and%
\begin{equation*}
\Gamma =(U-e^{-ik})^{2}+\gamma ^{2}-1.
\end{equation*}

From the elements of $S $ matrix one can easily obtain the transport properties of the system. In the case of only  injection from the left side, the elements $S_{12}$\ and $S_{22}$\ are the left
reflection and transmission amplitudes defined as $r_{L}=\frac{F_{b}^{-}}{F_{f}^{-}}$, and $t_{L}=\frac{F_{f}^{+}}{F_{f}^{-}}$, respectively. While in the case of only injection from the right side, the matrix elements $S_{11}$\ and $S_{21}$\ are the right transmission and reflection amplitudes defined as $t_{R}=\frac{F_{b}^{-}}{F_{b}^{+}}$, and $r_{R}=\frac{F_{f}^{+}}{F_{b}^{+}}$, respectively. So the $S $ matrix can be expressed as
\begin{equation}
S=\left(
\begin{array}{cc}
t_{R} & r_{L} \\
r_{R} & t_{L}
\end{array}
\right) . \label{SS}
\end{equation}
With the help of Eq. (\ref{Smatrix}), one can show that $t_{L}=t_{R}=t $ for a linear non-magnetic system \cite{ZinLin,Mostafa09}. For the $\mathcal{PT} $-symmetric system, one can prove the relation between transmission and reflection amplitudes as $r_{L}r_{R}^{\ast }= 1-\left\vert t\right\vert ^{2} $. Furthermore, defining the reflection and transmission coefficient as $R_{L,R}=\left\vert r_{L,R} \right\vert ^{2}$ and $T=\left\vert t\right\vert ^{2}$, we find that the generalized conservation relation $\sqrt{R_{L}R_{R}}=\left\vert T-1\right\vert $ holds in this $\mathcal{PT} $-symmetric non-Hermitian system, instead of the traditional relation $R+T=1$ for the Hermitian system \cite{LiGe12,YDChong11}. The transmission coefficient of this $\mathcal{PT} $-symmetric non-Hermitian model is found to be
\begin{eqnarray*}
&&T \\
&=&\frac{4-\omega ^{2}}{[U^{2}-U\omega +1]^{2}+(\gamma
^{2}-1)[2U^{2}-2U\omega +\omega ^{2}+\gamma ^{2}-3]}.
\end{eqnarray*}%
From the above equation, one can show that in the Hermitian case ($\gamma=0$), the resonant transport occurs when the income wave frequency equals to the potential strength. In this case, the central defects provide a path to guarantee a perfect transmission, i.e., $T=1 $ when $\omega=U $. This resonant state plays a significant role in the $\mathcal{PT} $ symmetry breaking.

According to the $S $-matrix theory, the breaking of $\mathcal{PT} $ symmetry may occur when we tune the strength of the non-Hermitian term in the scattering process. The system undergos a transition from the exact $\mathcal{PT} $ symmetric phase to the spontaneous $\mathcal{PT} $ symmetry breaking phase, which is distinguished by the eigenvalues of the $S $-matrix. In the exact $\mathcal{PT} $ symmetric phase, the eigenvectors of the $S $-matrix is also $\mathcal{PT} $ symmetric, and thus the eigenvalues are  unimodular. While in the $\mathcal{PT} $ symmetry breaking phase, the eigenvectors are not $\mathcal{PT} $ symmetric, and they transform into each other under the $\mathcal{PT} $ symmetry operator. Consequently, the two eigenvalues have reciprocal moduli, instead of being unimodular \cite{YDChong11}.

For the model (1), we can derive the analytical form of the eigenvalues of $S $-matrix. From Eq. (\ref{SS}), we have
\begin{eqnarray}
s_{1,2} &=&t\pm \sqrt{r_{L}r_{R}} . \nonumber
\end{eqnarray}
By using the property of the $S$ matrix, i.e., $S^*=S^{-1}$, we have $r_L=- \frac{t}{t^*}r^*_L$ and thus get
\begin{eqnarray}
s_{1,2}&=&t(1\pm \sqrt{\frac{T-1}{T}}) \nonumber \\
&=&t(1\pm \sqrt{g(\gamma ,U,\omega )\Delta })\label{s12},
\end{eqnarray}
where $g(\gamma ,U,\omega )$ and $\Delta $ are given by
\begin{equation*}
g(\gamma ,U,\omega )=-[\gamma ^{2}+U^{2}]/[4-\omega ^{2}]
\end{equation*}
and
\begin{equation}
\Delta =(\omega -U)^{2}+\frac{\gamma ^{2}(\gamma ^{2}+U^{2}-4)}{\gamma
^{2}+U^{2}},  \label{delta}
\end{equation}
respectively. While $g(\gamma ,U,\omega )$ is always negative for the permitted frequency $|\omega| \leq 2$, $\Delta $ serves as a discriminant to determine the moduli of the eigenvalues.
Eq. (\ref{s12}) indicates that the eigenvalues are unimodular for $T<1 $, i.e.,
\begin{equation}
\left\vert s_{1,2}\right\vert =1 ,
\end{equation}
whereas for $T>1 $, there exists
\begin{equation}
\left\vert s_{1}\right\vert =1/ \left\vert s_{2}\right\vert.
\end{equation}
So $T=1 $ can be taken as a criterion to signify the phase transition from $\mathcal{PT} $ symmetric phase to $\mathcal{PT} $ symmetry breaking phase, which is called the exceptional point (EP). The phase transition and EP can be also determined by the discriminant $\Delta $. One can easily show that the $\mathcal{PT} $ symmetric phase exists when $\Delta >0 $, while $\mathcal{PT} $ symmetry breaking phase exists when $\Delta <0 $. The criterion $T=1 $ is equivalent to the discriminant $\Delta=0 $.

From Eq. (\ref{delta}), we find that under the condition
\begin{eqnarray}
\gamma ^{2}+U^{2}-4<0, \label{sss}
\end{eqnarray}
there exist two EPs at
\begin{equation}
\omega _{\pm }=U\pm \sqrt{-\gamma ^{2}(\gamma ^{2}+U^{2}-4)/(\gamma ^{2}+U^{2}) } ,
\end{equation}
which is determined by $\Delta=0 $. For $\omega_{-}<\omega<\omega_{+} $, the system is in the $\mathcal{PT} $ symmetry breaking phase, whereas it is in the $\mathcal{PT} $ symmetry phase for $\omega<\omega_{-}$ or $\omega>\omega_{+} $. On the other hand, the system undergos no phase transition when $\gamma ^{2}+U^{2}-4 \geq 0 $ due to $\Delta >0 $, and thus the system is always in the $\mathcal{PT} $ symmetric phase. The positions of EPs $\omega_{\pm} $ also show that the symmetry breaking starts from the resonance state $\omega=U $ and spreads out to both sides when increasing the strength of non-Hermitian potential $\gamma $. This behavior indicates that the symmetry breaking is quite relevant to the resonant transport during the scattering propagation. Actually it is such the case because the $\mathcal{PT} $ symmetry breaking can also be signified by the transmission coefficient $T=1 $, and the system is in the perfect transmission in the resonant condition.
\begin{figure}[!htb]
\includegraphics[width=8.5cm]{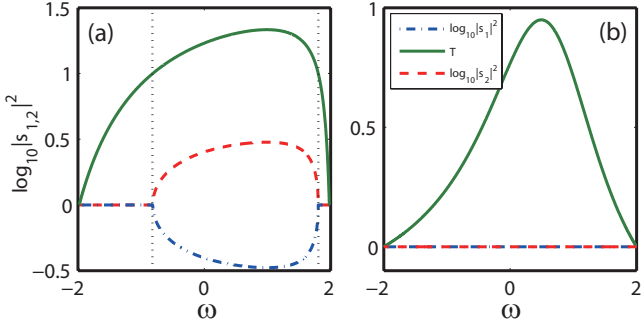}
\caption{(Color online) The eigenvalues of the $S $ matrix versus frequency $\omega $. $\log _{10}\left\vert s_{1}\right\vert ^{2} $  ($\log _{10}\left\vert s_{2}\right\vert ^{2} $) is presented by the blue dash-dotted (red dashed) line and the transmission coefficient is the green solid line. The black dotted lines are guided lines for eyes to indicate the positions of exceptional points. The parameters are $U=0.5, \gamma=0.5 $ in (a), and $\gamma=1.95 $ in (b).}
\end{figure}

To give concrete examples,  in Fig. 2 we show the numerical results of eigenvalues of the $S $ matrix as a function of $\omega $ for a certain $U=0.5 $. In Fig. 2(a) the strength of non-Hermitian potential is taken as $\gamma=0.5 $ which fulfils the condition given by Eq. (\ref{sss}). The results show that the system undergos two phase transitions at two EPs $\omega_{\pm}= 0.5 \pm \sqrt{1.75} $ as the incoming wave frequency $\omega $ changes. The system is in the $\mathcal{PT} $ symmetry breaking phase when the frequency $\omega_{-}<\omega<\omega_{+} $, which is signified by the reciprocal but non-unit moduli relation of $s_{1}$ and $s_{2}$. As shown in Fig. 2(a), blue dash-dotted and red dashed lines represent $\log _{10}\left\vert s_{1}\right\vert ^{2} $ and $\log _{10}\left\vert s_{2}\right\vert ^{2} $, respectively, and they are definitely zero when $\omega<\omega_{-}$ and $ \omega > \omega_{+} $, indicating that the system is in the $\mathcal{PT} $ symmetric phase. We also show the transmission coefficient $T$ denoted by the green solid line as a function of $\omega$ in the same figure. It is shown that $T<1$ in the $\mathcal{PT} $ symmetric phase, whereas $T>1$ in the broken $\mathcal{PT} $ symmetric phase. $T=1 $ defines the EPs just as $s_{1,2}$ defines them by the bifurcation of their modular. The dotted black lines mark the positions of two EPs in Fig. 2(a).
In Fig. 2(b), we show the eigenvalues of the $S $ matrix as a function of $\omega $ for a certain  $\gamma=1.95 $. In this case, the system is in the  $\mathcal{PT} $ symmetric phase in the whole $\omega $ region and no phase transition occurs, as $\gamma=1.95 $ obeys the condition $\gamma ^{2}+U^{2}-4 >0 $. Correspondingly, $T$ is always below unit in the whole $\omega $ region.

\section{CPA-Laser}
In the breaking $\mathcal{PT} $ symmetric phase of the non-Hermitian system, there is a unique absorption feature caused by the $\mathcal{PT} $ symmetry. As the simultaneous gain and loss are balanced embedded, the growth and decay modes are tightly linked each other in a unique manner. When the system permits self lasing at some frequency, it permits coherent perfect absorption (CPA) for a certain amplitude plane wave at the same frequency \cite{Longhi10}. The unique feature is known as CPA-Laser.

In this discrete model, we rewrite the scattering equation by the transfer matrix $M $,
\begin{equation}
\left(
\begin{array}{c}
F_{f}^{+} \\
F_{b}^{+}
\end{array}
\right) = M\left(
\begin{array}{c}
F_{f}^{-} \\
F_{b}^{-}
\end{array}
\right) \label{MM} .\\
\end{equation}
Supposing that the CPA-Laser occurs at frequency $\omega_{0} =2 \cos k_{0}$, the $\mathcal{PT} $ symmetry requires that the elements of the transfer matrix $M $ obey the relation $M_{11}(k_{0})=M_{22}^{\ast }(k_{0})=0 $ at this frequency $\omega_{0}$. In this discrete model, the elements of $M $ matrix are found to be
\begin{eqnarray*}
M_{11} &=&[1-(U-e^{ik})^{2}-\gamma ^{2}]e^{-ik}/(-2i\sin k) \\
M_{12} &=&[2U\cos k-U^{2}-\gamma ^{2}-2\gamma \sin k]e^{-ik}/(-2i\sin k) \\
M_{21} &=&[U^{2}+\gamma ^{2}-2U\cos k-2\gamma \sin k]e^{ik}/(-2i\sin k) \\
M_{22} &=&[(U-e^{ik})^{2}+\gamma ^{2}-1]e^{ik}/(-2i\sin k)  .
\end{eqnarray*}
The frequency for CPA-Laser in this model can be calculated following the relation discussed above, which leads to the restriction of the strength of non-Hermitian potential,
\begin{eqnarray}
\omega _{0} &=&2\cos k_{0}=2U, \\
\gamma  &=&\sqrt{2-U^{2}}. \label{CPA}
\end{eqnarray}
Despite that the $\mathcal{PT} $-symmetry breaking phase may exist when $\left\vert U\right\vert <2$, the CPA-Laser exists only under the condition $\left\vert U\right\vert < 1$ according to the above restriction, while the CPA-Laser does not exist when $1 \leq \left\vert U\right\vert <2$.

The CPA-Laser can be carved by an overall output coefficient $\Theta (k,\sigma )$ in $\mathcal{PT} $ symmetric system, defined as \cite{Longhi10}
\begin{eqnarray}
\Theta (k,\sigma)&=&\frac{\left\vert F_{b}^{-}\right\vert
^{2}+\left\vert F_{f}^{+}\right\vert ^{2}}{\left\vert F_{f}^{-}\right\vert
^{2}+\left\vert F_{b}^{+}\right\vert ^{2}} \nonumber \\
&=&\frac{\left\vert 1+\sigma M_{12}(k)\right\vert ^{2}+\left\vert \sigma -M_{21}(k)\right\vert ^{2}}
{(1+\left\vert \sigma \right\vert ^{2})\left\vert M_{22}(k)\right\vert ^{2}},
\end{eqnarray}
in which $\sigma =F_{b}^{+}/F_{f}^{-} $ represents the ratio of the injected signals from two sides, and in the second row we have used the property of the $M$-matrix, i.e., $M^*=M^{-1}$. It is noted that the output coefficient $\Theta (k,\sigma )$ is a useful sign to detect the simultaneous CPA-Laser, that is, the vanishing of coefficient $\Theta $ is the signature of coherent perfect absorption and $\Theta $ vanishes only for particular values of $\sigma=M_{21}(k_{0}) $, i.e., $\Theta (k_{0},\sigma=M_{21}(k_{0}))=0 $. For simplicity, this particular phenomenon can be understood from Eq. (\ref{MM}). Since the CPA requires that at the momentum $k_{0}$ there is $F_{f}^{-}, F_{b}^{+} \neq 0 $ and $F_{b}^{-}, F_{f}^{+} =0 $, so it's clear from the relation $F_{b}^{+}=M_{21}(k_{0})F_{f}^{-}+M_{22}(k_{0})F_{b}^{-} $ that $\sigma =F_{b}^{+}/F_{f}^{-} =M_{21}(k_{0}) $ is the equal criterion for CPA and Laser.
\begin{figure}[!htb]
\includegraphics[width=4cm]{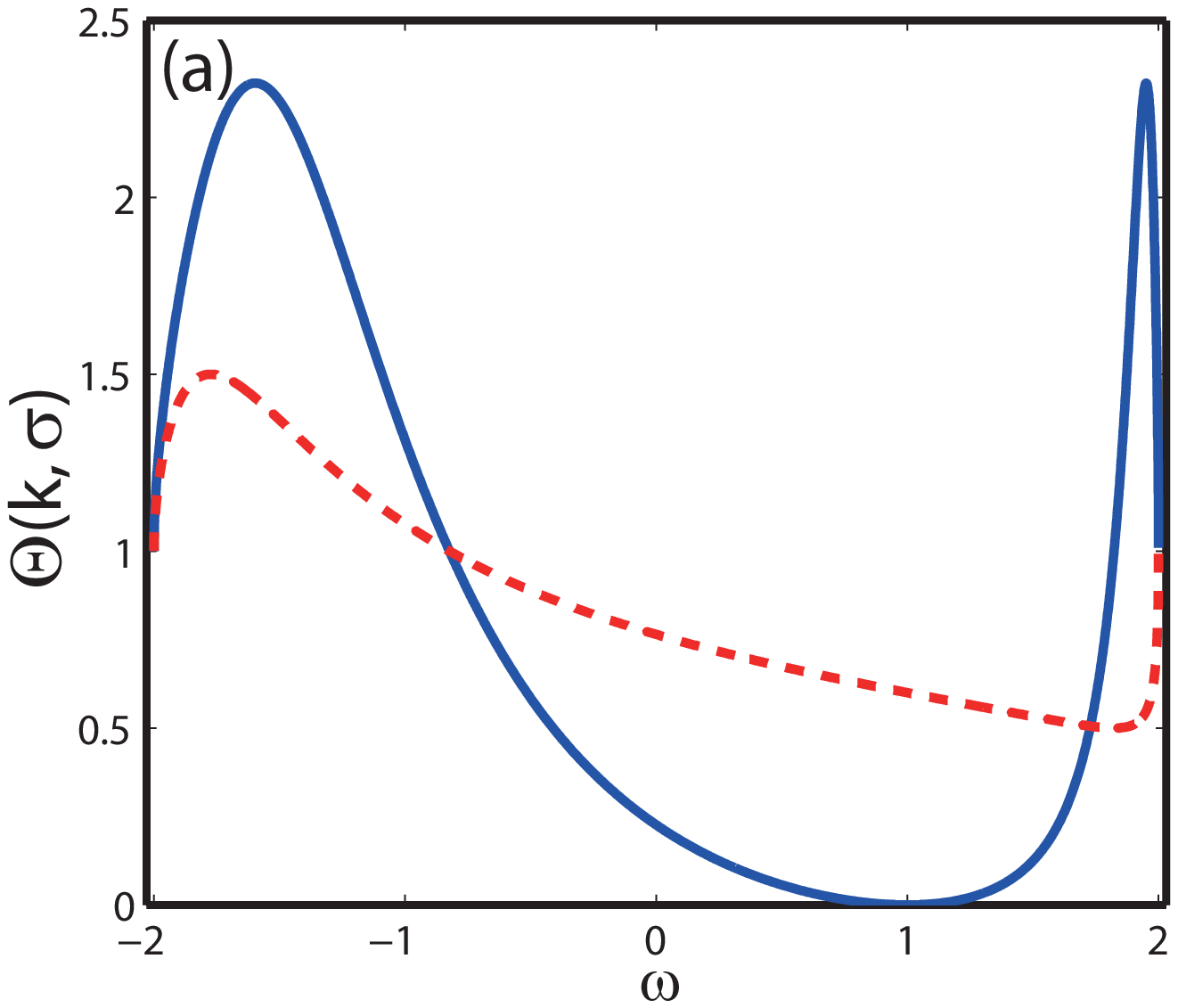} \\
\includegraphics[width=8.5cm]{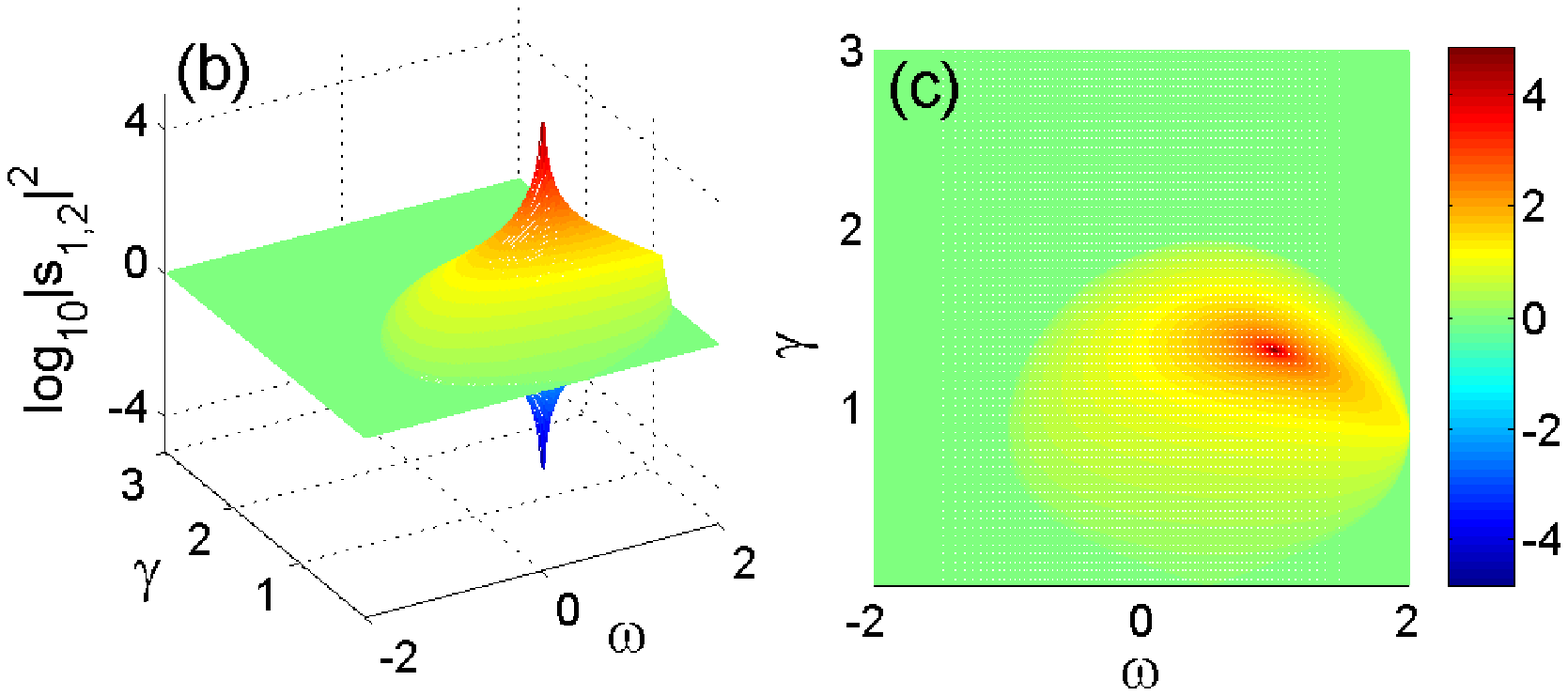}
\caption{(Color online) (a) Overall output coefficient $\Theta (k,\sigma )$ versus the frequency defining $\sigma = M_{21}(k)$. The parameter $U$ is taken as $U=0.5 $. The blue solid line stands for the case $\gamma= \sqrt{2-0.5^{2}}$ and the dashed red line for $\gamma=0.5 $. (b) $\log_{10}|s_{1,2}|^2$ versus $\gamma$ and $\omega$ for the system of $U=0.5$. (c) Aerial view of (b) from the top.}
\end{figure}

Fig. 3(a) shows the numerical results of $\Theta (k,\sigma )$ for $\sigma=M_{21}(k) $ as a function of frequency $\omega=2\cos k $ under the parameter $U=0.5 $. Two kinds of amplitude $\gamma$ are calculated. The blue solid line indicates that at the coherent $\gamma = \sqrt{2-0.5^2} $ required by Eq. (\ref{CPA}), the CPA-Laser occurs at frequency $\omega _{0} = 2\cos k_{0}=1$, which is signed by the vanish of $\Theta $. While for the other $\gamma $ marked by the red dashed line, there is no $k_{0}$ for CPA-Laser in the whole region as $\gamma $ can not follow the  Eq. (\ref{CPA}). There is another method to characterize the CPA-Laser by the eigenvalues of $S $ matrix. Fig. 3(b) shows that there is a sharp peak of the curve of $\log_{10}|s_{1,2}|^2 $ which indicates the appearance of a lasing mode at a certain $\gamma $ and $\omega $. Meanwhile there is also a deep peak indicating that it's an absorption mode at the same time. Fig. 3(c) gives an aerial view of Fig. 3(b) from the top and the results show that peaks appears at the frequency $\omega _{0} = 1, \gamma= \sqrt{2-0.5^2} $, which coincides with the results given by the output coefficient $\Theta $.

\section{Summary}
In summary, we have studied the symmetry-breaking
transition of the scattering process in a simple 1D $\mathcal{PT} $-symmetric non-Hermitian lattice model. With the help of the $S $-matrix method, we have analytically determined the exceptional points and given a discriminant to distinguish the $\mathcal{PT} $-symmetric phase and the breaking phase. In addition, we have shown that the unique feature of CPA-Laser can also be achieved in this simple model and obtained the condition to decide the appearance of CPA-Laser, which can be verified by the overall output coefficient $\Theta $ and the appearance of sharp peaks in the eigenvalues of $S $ matrix individually. This simple model can be easily implemented on experiments, such as the optical materials, the waveguide arrays,
and the acoustic sensor. Thus our work provides an additional way to explore the physical effect of the $\mathcal{PT}$-symmetric lattice system both in analytical and experimental fields and help us to understand the physical meaning of $\mathcal{PT}$ symmetry breaking in the lattice scattering problem.

\begin{acknowledgments}
This work is supported by NSF of China under Grants No. 11425419, No. 11374354 and No. 11421092. R. L. is supported by the NSFC under Grant No. 11274195 and the National Basic Research Program of China (973 Program) Grant No. 2011CB606405 and No. 2013CB922000.
\end{acknowledgments}

\end{document}